\documentclass{article}
\usepackage[T1]{fontenc}
\usepackage{wrapfig}
\usepackage[portuges,english]{babel}
\usepackage{amssymb,latexsym,amsmath,color,mathrsfs,ifsym,graphics,stmaryrd}
\usepackage{amsthm, amsfonts, mathrsfs}
\usepackage[colorlinks,linkcolor=blue,urlcolor=blue,citecolor=black,
plainpages=false,pdfpagelabels,breaklinks]{hyperref}
\usepackage[all]{xy}
\usepackage{graphicx,caption}
\usepackage[margin=2.1 cm]{geometry}

\newtheorem{theorem}{Theorem}[section]
\newtheorem{definition}[theorem]{Definition}

\newtheorem{postulate}[theorem]{Postulate}

\begin{document}

\title{{\sc On the Physical Untenability of the\\ Standard Notion of Quantum State}}

\author{{\sc Christian de Ronde}}
\date{}

\bibliographystyle{plain}
\maketitle

\begin{center}
\begin{small}
Philosophy Institute Dr. A. Korn, University of Buenos Aires - CONICET\\
Center Leo Apostel for Interdisciplinary Studies\\Foundations of the Exact Sciences - Vrije Universiteit Brussel\\
Institute of Engineering - National University Arturo Jauretche
\end{small}
\end{center}

\bigskip

\begin{abstract}
\noindent The notion of {\it quantum state} plays a fundamental role within the Standard account of Quantum Mechanics (SQM) as established by Dirac and von Neumann during 1930s and up to the present. In this work we expose the deep inconsistencies that exist within the multiple definitions of the notion of quantum state that are provided within this axiomatic formulation. As we will argue, these different inconsistent definitions continue to be ---even today--- uncritically confused within the mainstream physical and philosophical literature leading to self-contradictory statements and wrong conclusions. We end with a discussion regarding the untenability of this concept for any rational understanding of theoretical physics.  
\end{abstract}
\begin{small}

\end{small}

\bigskip

\bigskip

\bigskip

\bigskip

\bigskip

 \begin{flushright}
{\small {\it ``The classical concept of {\it state} becomes lost [in QM],\\ 
in that at most a well-chosen {\it half} of a complete set of\\ 
variables can be assigned definite numerical values.''}\\
\smallskip
Erwin Sch\"odinger} 
\end{flushright}

\section*{Introduction}

Today, it is commonly accepted within the physics community ---since Richard Feynman \cite{Feynman67} made it explicit to a packed audience of young undergraduate students in 1964--- that ``nobody understands quantum mechanics.'' This might seem very strange given that another common statement made by contemporary physicists, which contradicts the former, is that ``the theory describes a microscopic realm composed of elementary particles''. Of course, if the first statement is true the second one cannot be the case. This problematic situation regarding understanding is many times escaped by physicists who suddenly shift the attention to the fact the theory of quanta has been ---anyhow--- incredibly successful from a technological perspective. In this respect, it was the Manhattan project ---on which Feynman himself took part--- which exposed most clearly that understanding and technological development did not necessarily go hand in hand. Karl Popper would describe the radical shift that contributed to this new pragmatic account of physics during the late 1950s: 
\begin{quotation}
\noindent {\small ``Today the view of physical science founded by Osiander, Cardinal Bellarmino, and Bishop Berkeley, has won the battle without another shot being fired. Without any further debate over the philosophical issue, without producing any new argument, the {\it instrumentalist} view (as I shall call it) has become an accepted dogma. It may well now be called the `official view' of physical theory since it is accepted by most of our leading theorists of physics (although neither by Einstein nor by Schr\"odinger). And it has become part of the current teaching of physics.'' \cite[pp. 99-100]{Popper63}} 
\end{quotation}   
According to Popper, this new understanding of physics would become wildly accepted for two main reasons. Firstly, due to the influence of Bohr's complementarity approach in QM ---which he himself famously criticized \cite{vanStrien23}---, and secondly, due to the effectiveness of the {\it Manhattan Project}, exposed to the world with a ``big bang'' in August 6, 1945  \cite[pp. 101]{Popper63}. And still, regardless of its powerful technological applications, there are many dead ends that rise when attempting to explain or understand the Standard formulation of Quantum Mechanics (SQM) ---first presented by Dirac and von Neumann in the 1930s. 

But maybe the most powerful consequence of the instrumentalist account of physics, established after the IIWW \cite{Freire15}, is that the lack of understanding would not be regarded anymore as a problem. Physics would be understood as a discipline for predicting observations and ``solving problems''. Period. As a consequence, when faced to the many questions raised by critical students in physics classrooms, Professors would feel confident enough to shout to them: ``Shut up and calculate!'' There are many records of this common procedure that continues taking place even today (see for a detailed review \cite{deRonde25}). A good example is given by Lee Smolin who recalls: 
\begin{quotation} 
\noindent {\small ``When I learned physics in the 1970s, it was almost as if we were being taught to look down on people who thought about foundational problems. When we asked about the foundational issues in quantum theory, we were told that no one fully understood them but that concern with them was no longer part of science. The job was to take quantum mechanics as given and apply it to new problems. The spirit was pragmatic; `Shut up and calculate' was the mantra. People who couldn't let go of their misgivings over the meaning of quantum theory were regarded as losers who couldn't do the work.'' \cite[p. 312]{Smolin07}} 
\end{quotation}
In this same respect, Sean Carroll \cite{Carroll20} has recently described what happens ---even today--- to students that do not follow these basic instructions: ``Many people are bothered when they are students and they first hear [about SQM]. And when they ask questions they are told to shut up. And if they keep asking they are asked to leave the field of physics.'' But what are these difficulties with which students are faced when learning the standard Dirac-von Neumann axiomatic formulation of the theory of quanta? As we shall discuss in this work, the difficulties begin already with the first postulate of the theory which can be found in any textbook presentation of SQM. This first postulate, which focuses on the notion of {\it quantum state}, runs more or less as follows:

\begin{postulate}\label{postulate1} The possible {\it states} of a quantum system are represented by normalised vectors in some complex Hilbert space.
\end{postulate}

However, this is not the only definition that can be found within mainstream textbooks. There are ---at least--- four different definitions that are ---implicitly--- considered to be equivalent, consistent and meaningful. As we shall discuss in this work none of this is true. 

\smallskip

The paper is organized as follows. In section 1 we discuss four different definitions that can be found in any  mainstream SQM textbook. Section 2 discusses the irrepresentable rule that bridges the gap between quantum superpositions and observable measurement outcomes. In section 3 we discuss a widespread confusion present in the literature surrounding the notions of {\it pure state} and {\it quantum superposition} and, in section 4, we discuss what is considered to be {\it the same} through the notion of {\it physical state}. Finally, in section 5, we will argue that the physical untenability of the notion of quantum state implies a very deep problem for any rational theoretical representation of QM that attempts to scientifically discuss about an evolving state of affairs.



\section{The Orthodox Notion(s) of Quantum State}

Let us begin by considering the different mathematical definitions of the notion of quantum state that are presented within the mainstream physical literature. First of all, as presented in the first postulate, there is a widespread account of quantum states given in highly abstract mathematical terms, as referring to vectors in Hilbert space. This gives rise to the first definition: 
\begin{definition}[Abstract Vector State]\label{abstract} An abstract unit vector (i.e., with no reference to any basis), $\Psi$, in Hilbert space is a pure state. In terms of density operators $\rho$ is a pure state if it is a projector, namely, if Tr$(\rho^2) = 1$ or $\rho = \rho^2$. 
\end{definition}

As it is clear, this definition makes no reference whatsoever to any basis (i.e., reference frame). What we get is an abstract definition which, given that vectors are mathematically defined as invariant under coordinate changes, seemingly implies the reference to something which remains {\it the same}. This is of course the basic idea of the physical notion of {\it state} which, since Galileo, attempts to provide a mathematical representation of {\it the same} state of affairs even when considered from {\it different} basis dependent viewpoints. As we all know, in classical mechanics it is the Galilean transformations which provide consistency to the values of properties described from different reference frames. This means that the different coordinate system representations of a rabbit running through a field, when being described by two different observers, one in a train and another one in the train-station, will be consistent. In short, it is the Galilean transformations, by providing a consistent {\it operational-invariant} translation between the different reference-frame dependent representations, which allow to talk about the rabbit from different viewpoints. Any observer can place himself within the viewpoint of any other observer and compute all the values of properties characterizing the system under investigation from that particular perspective. All perspectives are then equivalent. And that is what, in fact, allows to talk consistently about the same  `state of affairs' from different perspectives or viewpoints. 

At this point, it is important to stress that the abstract vector state definition \ref{abstract} does not make reference to the operational-invariance we just mentioned. Instead, the invariance that is implied within this first definition is purely mathematical, completely abstract (e.g. \cite{BZ17, Hall13}). But this does not mean that we lack, within the mainstream literature, an operational definition of a quantum state. On the contrary, we also find, in textbooks that prefer a more ``pragmatic'' approach to SQM  ---such as the very good book by Asher Peres {\it Concepts and Methods} \cite{Peres02} or the famous book by Nielsen and Chuang \cite{NielsenChuang10}---, another widespread account of the notion of pure state given in operational terms, through a distinction between the different type of predictions implied by quantum states when represented in different bases. This second definition runs as follows: 

\begin{definition}[Operational Pure State]\label{operational} Given a quantum system in the state $\Psi$, there exists an experimental situation linked to a specific basis in which the test of it will yield with certainty (probability = 1) its related outcome. That particular basis is the one in which the abstract vector  $\Psi$ is written as a single term ket $|\psi \rangle$.  
\end{definition}

Contrary to the previous one, this second definition of a quantum pure state in SQM is basis dependent, it depends on the specific basis in which the abstract vector will be represented as a single term {\it ket}. It is only in {\it that} particular basis that we will be {\it certain} about the outcome when measuring the state. Let us remark that the certainty is restricted to a prediction with probability = 1. Now, what is important to notice is that this definition is essentially contextual and thus non-invariant, this notion of state depends on a specific ``preferred basis''. Something which ---of course--- does not exist in the context of any other classical or relativistic theory where all reference frame dependent representations are considered to be physically equivalent. 

As it is well known, apart from these operational {\it pure} or {\it maximal} states, which are those that provide (binary) certainty about the prediction of measurement outcomes in a particular basis, we also have states that can be in a {\it quantum superposition of states}. As it was stressed by Dirac himself: ``One of the most fundamental and most drastic laws of [QM] is the {\it Principle of Superposition}.'' More specifically:  
\begin{quotation}
\noindent {\small ``[The Superposition Principle] requires us to assume that between these states there exist peculiar relationships such that whenever the system is definitely in one state we can consider it as being partly in each of two or more other states. The original state must be regarded as the result of a kind of {\it superposition} of the two or more new states, in a way that cannot be conceived on classical ideas. Any state may be considered as the result of a superposition of two or more other states, and indeed in an infinite number of ways. Conversely any two or more states may be superposed to give a new state. The procedure of expressing a state as the result of superposition of a number of other states is a mathematical procedure that is always permissible, independent of any reference to physical conditions, like the procedure of resolving a wave into Fourier components. Whether it is useful in a particular case, though, depends on the special physical conditions of the problem under consideration.'' \cite[p. 12]{Dirac74}}
\end{quotation}

The superposition principle in QM implies that the linear combination of different states which are solutions to the equation of motion can be also considered as a legitimate physical state. These states are, in fact, the most common states within the vectorial formulation of the theory. Obviously, almost all bases lead to a representation of states in terms of quantum superpositions, not in terms of {\it maximal states}. 

\begin{definition}[Superposition State]\label{superposition} Given a quantum system in the (abstract) state $\Psi$, there are many different bases in which the vector can be represented as a linear combination of different {\it kets}.  
$$\sum_{i} c_{i} \  |\phi_i \rangle $$
In this case, each possible result, linked to a particular state $|\phi_k \rangle$, will lead to a probabilistic {\it uncertain} measurement outcome (computed through the Born rule $| c_k|^2$). Given that the theory does not predict the actual measurement outcome, the (binary) certainty is lost. 
\end{definition}

It is then important to understand that, all  quantum states represented in a particular basis, maximal or not, make reference to a particular measurement context, and consequently, to a specific state of affairs. Thus, even though one might depart from the same abstract vector state, $\Psi$, each reference frame basis-representation of that state will make reference to a {\it different} state of affairs. This relativist idea, which is kernel to SQM, goes of course back to Bohr's notion of complementarity according to which a quantum state of affairs can be only considered once the measurement context is explicitly determined \cite{Bohr35}. In short, the idea is that quantum reality is essentially contextual, namely, relative to the choice of the basis or experimental situation. Given the complementary nature of quantum states, a state considered in one particular basis will be {\it different} to that same abstract vector state considered in another different basis. This idea, which remains part of the common wisdom in mainstream SQM, implies that the {\it quantum state} is always {\it relative} to a specific chosen context, a Complete Set of Commuting Observables or simply put, a basis of that Hilbert space. 

Two remarks go in order regarding the contextuality of quantum states. The first one is that the proof that there is no operational-invariance between the different reference frames dependent representations of the same abstract quantum state is well known within the foundational and philosophical literature and was famously demonstrated by a theorem due to Simon Kochen and Ernst Specker in 1967 \cite{KS}. Unfortunately, the fact that there is no {\it global binary valuation} of the projection operators of a quantum state has been taken to imply ---following Bohr, Dirac and von Neumann--- that the choice of a basis is a necessary precondition to consider a physical state. It should be stressed, however, that this is not necessarily the case. In fact, when going back to Heisenberg's original intensive formulation of QM the operational-invariance is present right from the start and, as it has been demonstrated in \cite{deRondeMassri21}, a {\it global intensive valuation} is always possible. The second remark is that this relativist account of quantum states, which is taken for granted by physicists within SQM, has been extensively discussed in philosophical journals where a countless number of ``interpretations'' have been proposed. We might mention just a few of the most popular interpretations which have taken as a standpoint the contextual or relativist nature of quantum states: the ``relative state'' interpretation by Hugh Everett \cite{Everett57}, the ``relational interpretation'' presented by Carlo Rovelli \cite{Rovelli96}, the ``perspectival modal interpretation'' discussed by Dennis Dieks \cite{Dieks22} and the ``contextual interpretation'' proposed by Philippe Grangier \cite{Grangier02} are all good examples.\footnote{It is important to stress, however, that the working quantum physicists who are trained in an instrumentalist fashion do not care at all about this ``philosophical debate'' which, as a matter of fact, and due to its many problems, has been recently characterized by \'Adan Cabello ---a prestigious physicist in the field of quantum information--- as a ``map of madness'' \cite{Cabello17}.}

And still, the definitions we provided above do not exhaust the consideration of quantum states. Last but not least, we also find within the mainstream literature another definition of state that was conceived, in empirical-positivist terms, through the direct reference to observations. This reference was imposed by no one else then Paul Dirac \cite[p. 10]{Dirac74} himself who would argue explicitly that: ``the main object of physical science is not the provision of pictures, but the formulation of laws governing phenomena and the application of these laws to the discovery of phenomena. If a picture exists, so much the better; but whether a picture exists of not is a matter of only secondary importance.'' In short, that it is ``important to remember that science is concerned only with observable things''. Following these considerations the quantum state was also defined ---in purely empirical terms--- as linked to the actual observation of a particular outcome, after a measurement had been actually performed on the (pre-measured) quantum state. 

\begin{definition}[Empirical State]\label{empirical} After a measurement takes place on a quantum state, we obtain a single measurement result which can be understood as the new quantum state of the system (after measurement).   
\end{definition}

Of course, the only quantum state in which there will be a coincidence between the state before and after the measurement is the {\it maximal state}. The problem is that when considering quantum superpositions of states ---which, in fact, are obviously the most common states within the vectorial formulation--- such coincidence does not exist. Thus, in order to bridge the gap between the intensive patterns described by quantum superpositions and the binary `clicks' observed in detectors, Dirac would introduce, as an axiom of the theory itself, the famous ``collapse of the quantum wave function''.  Of course, such introduction of a non-linear evolution within a linear mathematical formalism creates another serious inconsistency to which we will now turn our attention.

\section{From Superposed States to Empirical States}

Even though Dirac stressed the importance of quantum superpositions within the theory of quanta, he would also argue that the essential goal of any physical theory is to account for ``observations''; by which he meant ``single clicks in detectors''. Of course, there is a huge gap between the mathematical representation provided by quantum superpositions in terms of intensive patterns and single measurement outcomes considered ---since Max Born's interpretation of the quantum wave function--- as the consequence of microscopic particles. While the first provides a non-classical intensive representation of a state of affairs, the latter representation is essentially binary. As it is well known, in order to bridge the gap between superpositions and (binary) observations Dirac would impose the existence of a non-linear evolution between them, a ``collapse'' that would allow to transform superpositions into single outcomes. This non-linear evolution, contradicting the original mathematical formalism, would become introduced within his axiomatic formulation, in an {\it ad hoc} manner, as a main {\it postulate} of the theory itself. This would become known in the mainstream literature as the ``projection'', ``collapse'' or ``measurement'' postulate. 

It is important to stress that the reference of SQM to single measurement outcomes, established since the work of Dirac and von Neumann as an unquestionable dogma, has remained an essential element present in contemporary physics and philosophy of physics. In this respect, two remarks go in order. The first is with respect to the distinction between collapses and the Born rule. It has been argued that: 
\begin{quotation}
\noindent {\small ``The projection postulate appears in Dirac (1930) and von Neumann (1955), the first two codifications of the axioms of QM. It continues to be widespread, though not universal, in first courses on QM to this day: an unscientific perusal of my shelf reveals that collapse is included in about half of the books there that present QM from scratch. (The Born rule, of course, appears in all of them.)'' \cite[p. 288]{Wallace19}}
\end{quotation}
This comment might lead to the wrong conclusion that the projection postulate is not always considered by physicists. But this not correct. The 1926 interpretation by Max Born of the quantum wave function already contains an implicit ``proto-collapse'' completely analogous to that of Dirac which gives rise to the measurement problem (see for a detailed analysis \cite{Landsman17}). Just in the same way as Dirac's collapse, Born's rule had already created an irrepresentable bridge between the intensive representation provided by quantum wave functions and the single `clicks' observed in detectors, interpreted ---in turn---  as the ``self-evident'' effect of ``quantum particles'' themselves. In his paper titled {\it On the Quantum Mechanics of Collision Processes}, Born interpreted $\psi (x)$ as encoding a probability density function for an elementary particle to be found at a given region. The wave function, a complex-valued function, had physical meaning only in terms of square modulus, $|\psi(x)|^2 = \psi (x)^* \psi (x)$. If $|\psi (x)|^2$ has a finite integral over the whole of three-dimensional space, then it is possible to choose a normalizing constant and the probability that a particle is within a particular region V is the integral over V of $|\psi (x)|^2$. In this way, Born's rule was imposing a radical shift in the reference of quantum probability from intensive values ---which Heisenberg had originally considered as the observational standpoint of QM--- to single measurement outcomes. As remarked by Born \cite{Born26}: ``Schr\"{o}dinger's quantum mechanics [therefore] gives quite a definite answer to the question of the effect of the collision; but there is no question of any causal description. One gets no answer to the question, `what is the state after the collision' but only to the question, `how probable is a specified outcome of the collision'.'' Of course, even though the reference to particles and collisions was explicit within Born's narrative there was no theoretical representation of any of them. 
 
The second remark is that, even though within the philosophical debate there is a common distinction between ``collapse'' and ``non-collapse'' interpretations of SQM, both of these groups take for granted the reference of the theory to single measurement outcomes. In fact, this reference is assumed as a main standpoint in all interpretations of SQM and, consequently, the projection postulate remains untouched. The only difference is that while in the case of ``collapse interpretations'' it is believed that the postulate implies the existence of ``a real physical process'', within the second ``non-collapse'' group, the projection postulate is reinterpreted in a non-realist fashion. For example, in Dieks' modal interpretation it is not understood as a ``real collapse'' but ---instead--- as some kind of ``pragmatic'' or ``instrumental'' rule that is required in order to obtain the observation of single `clicks' in detectors. Following the van Fraassen's semantic account of theories Dieks \cite{Dieks91} argues that: ``an uninterpreted theory is identified with the class of its models, in the sense of abstract model theory. [...] To make an empirical theory of it, we have to indicate how empirical data can be embedded in the models.'' In particular, in order ``[t]o obtain quantum mechanics as an empirical theory, we have to specify the links with observation, by means of `interpretation rules'.'' This is presented by Dieks in the following terms:
\begin{quotation}
\noindent {\small ``Consider a state vector representing a composite system, consisting of
an object system and the remainder of the total system. The total state
vector will almost always have one unique bi-orthonormal decomposition:
\begin{center}
$| \Psi \rangle = \sum c_k  |\psi_k \rangle |R_k \rangle $ \ \ \ \ \ \ \ \ \ \ \ \  (2)
\end{center}
where the $|\psi_k \rangle$ refer to the object system and the $|R_k \rangle$ to the rest of the
system; $\langle \psi_i | \psi_j \rangle$ and $\langle R_i | R_j \rangle = \delta_{ij}$. I now propose the following {\it semantical
rule:} {\bf As soon as there is a unique decomposition of the form (2), the
partial system represented by the $|\psi_k \rangle$, taken by itself, can be described as
possessing one of the values of the physical quantity corresponding to the set $\{ |\psi_k \rangle \}$. The probabilities for the various possibilities to be realized are given by $| c_k|^2$.}'' \cite[p. 1406]{Dieks89}}
\end{quotation}
This ``semantic'' or ``interpretational'' rule is nothing essentially different from the measurement rule or projection postulate, it simply denies the relation between the rule ---namely, the mathematical formalism--- and physical reality. But one might then ask why only a restricted part of the mathematical formalism needs to be interpreted realistically and another part does not? in fact, departing from realism, a similar solution was provided by Hugh Everett in his relative state interpretation where the ``collapse'' is reinterpreted as an ``epistemological branching''. In his PhD dissertation, taking as a standpoint Wigner's friend paradox Everett would attempt to avoid collapses ---like Dieks--- without abandoning the positivist requirement to refer to actual observations. His strategy was basically to give up on metaphysical narratives and understand SQM in purely epistemological terms. In his own words \cite[p. 253]{BarrettByrne}: ``To me, any physical theory is a logical construct (model), consisting of symbols and rules for their manipulation, some of whose elements are associated with elements of the perceived world.'' In this way, following the positivist program, Everett's account of SQM would secure the {\it reference} to `clicks' but only as being {\it relative} to agents. In this respect, he would also stress: ``There can be no question of which theory is `true' or `real' ---the best that one can do is reject those theories which are not isomorphic to sense experience.'' Once SQM was understood by Everett as making reference to the relative observations made by agents, the `collapse' seemed to have finally disappeared. And yet, the measurement rule (or projection postulate) had not. Instead, it had changed its meaning from an un-observable ``measurement process'' to an even stranger ``branching process'' also dependent on subjects which had to be understood in purely operational terms \cite{BarrettByrne}. One decade later, during the early 1970s, quite regardless of Everett's intentions, Bryce DeWitt's and Neill Graham would attempt to conceive this branching process now in ontological terms. Turning Everett's relativist epistemological account of QM upside-down, DeWitt and Graham would claim that the quantum measurement interaction described by the branching process was actually responsible for creating a multiplicity of ``real worlds''.\footnote{This explains why Jeff Barrett, who had access to Everett's original notes, found written next to the passage where DeWitt presented Graham's many worlds clarification of Everett's own ideas the word ``bullshit''. See \cite[364-6]{BarrettByrne} for scans of  Everett's comments.} In DeWitts' and Graham's many worlds reformulation the ``branching process'' became once again ---just like the ``collapse''--- a real physical process, one which was able to create from a single measurement a myriad of (truly real) parallel worlds. Once again, nothing of this incredible ``creationist process'' of universes was described by the theory.\footnote{In this respect, it is very interesting to notice that we could think of the ``branching process'' as the mirror image of the ``collapse process'' none of which is addressed nor explained by QM. While the collapse turns the superposition into only one its terms, the branching goes from one single measurement into a multiverse of parallel worlds.}

\section{Are Quantum Superpositions Pure States?}

Another confusion that appears in the mainstream literature, which highlights the problems we have been addressing can be exposed through the following very simple question: Are quantum superpositions pure states? It is easy to show that, depending on the choice of the definition, one can answer this question positively and also negatively. Let us see how this works.  

In mathematical terms we can always write the following equivalence relation which appears in most papers which attempt to explicitly address quantum superpositions: 
$$|\psi \rangle = \sum_{i} c_{i} \  |\phi_i \rangle $$ 
\noindent Here we are using the first definition \ref{abstract} of a quantum state (section 1) in order to present a mathematical equivalence. So, of course we need to answer the question positively. All these states are representations of the same vector and thus must be regarded as the same quantum superposition. 

As discussed above, even though this is a mathematical equivalence, it is not an operational equivalence because ---due to Kochen-Specker theorem--- there is no invariant global valuation when considering different bases. In fact, given the mathematical equivalence, it should be strange to a mathematically trained reader to always find the same equivalence relation in papers which discuss about quantum superpositions. Why not, for example, write the following equivalence: 
$$ \sum_{i} c_{i} \  |\phi_i \rangle = \sum_{i} d_{i} \  |\gamma_i \rangle $$ 
As representations of the same abstract state $\Psi$, $|\psi \rangle$, $\sum_{i} c_{i} \  |\phi_i \rangle$ and $\sum_{i} d_{i} \  |\gamma_i \rangle$ should be all regarded as completely equivalent mathematical representations of the same state (see definition \ref{abstract}). Right? But this is clearly not the case in SQM where we find a fundamentality of states written as a single term {\it ket} which is not present in those states constituted by a sum of different states. This is of course due to their operational content. While states with more than one term like  $\sum_{i} c_{i} \  |\phi_i \rangle$ and $\sum_{i} d_{i} \  |\gamma_i \rangle$ do not provide certainty with respect to measurement predictions, the state $|\psi \rangle$ ---which is clearly not a superposition of states--- does provide (according to definition \ref{operational}) complete certainty. This is in fact one of the widespread definitions of a {\it pure state} as a state that provides maximal knowledge about its measurement result. On the contrary, superposed states like $\sum_{i} c_{i} \  |\phi_i \rangle$ and $\sum_{i} d_{i} \  |\gamma_i \rangle$ are known to be uncertain. Consequently, following this second line of reasoning, the answer must be negative. Quantum superpositions are not pure states. 

To conclude, depending on the choice of the many definitions that one can find within SQM (section 1), one can choose at will to answer the question in positive or negative terms. This is of course consequence of the  (in-)equivalence between the multiple definitions that are used within the literature.

\section{Quantum States: Different or The Same?}

Physics was born in Greece during the 5th and 4th century B.C. as related to the problem of movement, namely, how to account for the multiplicity of phenomena in terms of a unity, how to account for identity within difference, how to conceive the many within the one. As Wolfgang Pauli would explain to a young Heisenberg:  
\begin{quotation} 
\noindent {\small  ``[...] knowledge cannot be gained by understanding an isolated phenomenon or a single group of phenomena, even if one discovers some order in them. It comes from the recognition that a wealth of experiential facts are interconnected and can therefore be reduced to a common principle. [...] `Understanding' probably means nothing more than having whatever ideas and concepts are needed to recognize that a great many different phenomena are part of coherent whole. Our mind becomes less puzzled once we have recognized that a special, apparently confused situation is merely a special case of something wider, that as a result it can be formulated much more simply. The reduction of a colorful variety of phenomena to a general and simple principle, or, as the Greeks would have put it, the reduction of the many to the one, is precisely what we mean by `understanding'. The ability to predict is often the consequence of understanding, of having the right concepts, but is not identical with `understanding'.'' \cite[p. 63]{Heis71}} 
\end{quotation}

Searching to account for {\it the same} within {\it change} physics reached with Aristotle a conceptual/categorical definition of an {\it actual entity} in terms of the principles of existence, non-contradiction and identity, solving in this way the ancient problem of movement. Two millennia later, in modernity, this representation would reach its highest peak with the creation of classical mechanics. Essential to this development was, firstly, the work of Galileo regarding the equivalence between reference frames and then, secondly, the possibility to mathematically represent continuous space and time through infinitesimal calculus. In this context, the notion of {\it physical state} was developed in terms of Galilean transformations, as the moment of unity that allowed to account invariantly and consistently for the values of properties of the system independently of reference frames. It is this invariant definition which allowed to talk in classical mechanics about {\it the same} independently of any viewpoint. In this way, all observers, even though could see things differently (depending on their perspective), would nonetheless posses a translation from one perspective to the other allowing to consistently account for a common referent. 

On the very contrary, in the context of SQM, first Bohr and then Dirac would subvert completely the notion of (quantum) state, turning it explicitly dependent on experimental perspectives and reference frames. Bohr \cite[p. 7]{WZ} would replace the systematic definition of state by a commonsensical reference to classical objects arguing that: ``[...] the unambiguous interpretation of any measurement must be essentially framed in terms of classical physical theories, and we may say that in this sense the language of Newton and Maxwell will remain the language of physicists for all time''. Furthermore, as he would warn his followers, ``it would be a misconception to believe that the difficulties of the atomic theory may be evaded by eventually replacing the concepts of classical physics by new conceptual forms.'' In this context, quantum objects would come to require complementary ---incompatible--- classical representations such as that of `wave' and `particle' that would explicitly depend on the choice of a particular experimental context (or, in mathematical terms, a basis given by a complete set of commuting observables). Destroying objectivity (i.e., the categorical construction of a {\it moment of unity} capable to account consistently for a multiplicity of different phenomena) Bohr's notion of complementarity would then impose an inconsistent contextual description where the pre-requisite to discuss about quantum objects would be the actual choice or effectuation of an experimental arrangement. As he would famously argue in his reply to the EPR paper: 
\begin{quotation}
\noindent {\small ``it is only the mutual exclusion of any two experimental procedures, permitting the unambiguous definition of complementary physical quantities, which provides room for new physical laws, the coexistence of which might at first sight appear irreconcilable with the basic principles of science. It is just this entirely new situation as regards the description of physical phenomena, that the  notion of complementarity aims at characterizing.'' \cite[p. 700]{Bohr35}}
\end{quotation}


As we discussed above, following the Bohrian-positivist premises, Dirac would then impose a radically new contextual definition of the notion of {\it state} which has remained untouched up to the present. Each (quantum) {\it state}, mathematically represented by ---what he himself called--- a {\it ket}, $|x\rangle$, would be linked not only to a single measurement observation (i.e., a `click' in a detector) but also to a specific reference frame (or basis). This means that a {\it ket} is not an {\it abstract vector}, $x$, but one already represented in a specific basis. Confusing his own definitions, Dirac would also argue that {\it the same system} could be represented in terms of {\it different states}, relative to the different bases: 
\begin{quotation}
\noindent {\small ``[E]ach state of a dynamical system at a particular time corresponds to a ket vector, the correspondence being such that if a state results from the superposition of certain other states, its corresponding ket vector is expressible linearity in terms of the corresponding ket vectors of the other states, and conversely. Thus the state $R$ results from a superposition of the states $A$ and $B$ when the corresponding ket vectors are connected by $ | R \rangle\  = c_1 | A \rangle\ + c_2 | B \rangle\ $.'' \cite[p. 16]{Dirac74}}
\end{quotation}

Thus, depending on the chosen basis (or reference frame), the representation of {\it the same} state would be given in terms of {\it different} states. This means that in Dirac's formulation bases would not represent {\it the same state}, but instead would impose the distinction between {\it different states}. Each reference frame (or basis) would make reference to a particular state ---in general--- composed of many (superposed) states. What is essential to recognize is that the relativism involved here differs significantly from that imposed by classical physics and ---even--- relativity theory when considering different reference frames. While in the case of classical physics and special relativity the relative values with respect to different reference frames can be consistently and invariantly considered from within a global transformation, allowing to define a common reference ---namely, a common account of {\it the same} state of affairs---, in the Bohr-Dirac ``standard'' account of QM each different reference frame describes a {\it different} state of affairs, essentially incompatible with the others. And in this case, it is ``the preferred basis'', the one that is actually chosen or measured, which provides a restricted form of contextual representation (see \cite{BK22, Hemmo22, Zurek93}). Thus, while in the first case what is relative to the different reference frames is not really decisive, since we are capable to consistently consider the different frame-dependent representations altogether, in the latter case we might talk about the introduction of a fundamental {\it perspectival relativism}, where the impossibility of a global consistent account is established not as a fundamental problem but ---instead--- as a feature of the theory itself. This implies the most radical shift in the history of physics from objectivity and invariance to complementarity and contextuality, destroying not only the notion of physical state but also the possibility of a dynamical representation.

The inconsistencies we have pointed should already be clear to the attentive reader (see for a more detailed analysis \cite{deRondeMassri22}). However, let us make even more explicit the contradiction that is reached when applying these inconsistent definitions (see Table 1). 
\medskip
\begin{center}
\begin{tabular}{c|l}
Mathematical symbol & Physical interpretation \\ 
\hline
$\Psi$              & Abstract vector state \\
$|\uparrow_x\rangle$                    &  A one term state representing $\Psi$ in the basis $\{ |\uparrow_x\rangle, |\downarrow_x\rangle\}$. \\ 
$a|\uparrow_y\rangle+b|\downarrow_y\rangle$ &  A superposition state as a sum of different states \\ & when representing $\Psi$ in the basis $\{ |\uparrow_y\rangle, |\downarrow_y\rangle\}$.
\end{tabular}
\captionof{table}{Mathematical and physical accounts of vectors and states.}
\end{center}
\medskip
From a mathematical perspective it is clear that in the first column we have different representations of the same state vector. However, from a physical perspective, things are quite different. Even though there is no contradiction in mathematical terms, from a physical perspective {\bf a state that is certain cannot be uncertain}. The state $|\uparrow_x\rangle$ is {\it certain} when measured in the basis $\{ |\uparrow_x\rangle, |\downarrow_x\rangle\}$, but {\it the same} abstract vector (i.e., the same state) is also given by the state $a|\uparrow_y\rangle+b|\downarrow_y\rangle$ which is {\it uncertain} when measured in the basis $\{ |\uparrow_y\rangle, |\downarrow_y\rangle\}$. This is a clear inconsistency: the same state cannot be certain and uncertain at the same time. The inconsistency rises because {\it the same} abstract state vector $\Psi$ is related through the different basis dependent representations of the vector (i.e., $|\uparrow_x\rangle$  and  $a|\uparrow_y\rangle+b|\downarrow_y\rangle$) to different physical states of affairs. Or in other words, a {\it mathematical equivalence} is not necessarily a {\it physical equivalence}. The term {\it state} is then used to refer to different incompatible situations which are then also considered to be the same in abstract mathematical terms. As a consequence, the states $|\uparrow_x\rangle$,   $a|\uparrow_y\rangle+b|\downarrow_y\rangle$, $|\uparrow_y\rangle$ and $|\downarrow_y\rangle$ are different and the same. Justified by the self-apologetic explanation that ``QM is weird'', this widespread confusion has led to the uncritical acceptance within the mainstream physical and philosophical literature of an essentially inconsistent discourse. This contradiction can be clearly understood through the definitions provided in section 1. According to definition \ref{abstract}, all bases provide a mathematical representation of the same state $\Psi$. However, according to definition \ref{operational}, operational pure states are different from the superposed states described in definition \ref{superposition}. How is it possible that the same state, $\Psi$, gives rise to different mutually incompatible basis dependent states? Basically, because we are using implicitly two different references of what must be considered ``the same''. While the first is a mathematical purely abstract account of unity, the second is an operational definition grounded on a binary certainty that is completely alien to the theory. Even though there is a mathematical transformation from one basis to another of the same abstract vector state, there will be no consistency of the values of properties between these different mathematical representations. As exposed by the Kochen-Specker theorem \cite{KS}, even though there is a mathematical abstract-invariance there is no physical operational-invariance of the values of projection operators (see section 2). 

\section*{Conclusion}

As we have shown in this work, the notion of {\it state} within SQM is grounded in a series of inconsistent and incompatible definitions which are commonly confused in the physical and philosophical literature. As it stands, this untenable concept will certainly obstruct any physical or technological development which attempts to provide any meaningful rational account of quantum phenomena. Instead of uncritically accepting these inconsistencies, it seems it is time for a radical reconsideration of the notion of quantum state. One which is capable to replace this inconsistent reference by a  meaningful physical discourse that connects consistently an operational invariant formalism with truly objective concepts.

\section*{Acknowledgements} 

The author state that there is no conflict of interest.

\end{document}